# Discontinuity in the Brightness of the Twilight Sky at Different Wavelengths

Nawar S., Morcos A. B., Tadross A. L., Mikhail J. S.

*National Research Institute of Astronomy and Geophysics, Helwan, Cairo, Egypt*

## Abstract

A search for discontinuity in the sky twilight brightness at different wavelengths and different solar depressions for altitudes 0, 5, 10, 30, 50 & 70 degrees is done. It is found that the logarithmic of difference in the brightness log ($I_1$-$I_2$) of two similar patches lying at solar and anti-solar verticals are not suffering of any discontinuity, when plotted versus sun's depression. Whatever, the ratio $I_1/I_2$ shows that there is discontinuity in the curves at altitudes less than or equal 30 degrees. While for altitudes 50 & 70 degrees slight patches of discontinuities have been detected. The phenomenon of discontinuities may be referred to the changes in the sky twilight brightness due to the effect of the Earth's shadow on diffusing and luminescent layers in the upper atmosphere.

## Introduction

Different investigators have examined the discontinuities in the brightness of the sky twilight. Grandmontagne (1941) has found that the curve of logarithm of brightness against the solar depression shows a discontinuity in the brightness of the sky twilight. These discontinuities were found to repeat themselves on different days. Gauzit and Grandmontagne (1942) have explained this phenomenon as actual discontinuities of the density gradient at upper atmosphere. He confirmed his suggestion by using rocket data of upper atmospheric temperature. Vacouleurs (1951) has confirmed the existence of the discontinuities at solar depressions 6.2, 8.7, 9.1 & 12.1 degrees. He suggested that this effect is a result of luminescent layers in the upper atmosphere.

Karandikar (1955) studied a type of discontinuities enhanced by observing two similar patches of the zenith sky at the same solar meridian. He has found from his observations of a single spot near zenith that there is no evidence of discontinuity in the brightness of astronomical twilight.

In present work we are going to study the discontinuities in the twilight observations using the same principles of Karandikar's method (1955). In the first section we will explain shortly the mean ideas of karandikar's method. The results will be discussed in the second section. At the end our conclusion will be given.

## Principles of Method

The principle idea of the method of studying the discontinuities in observations of astronomical twilight is depending on observing two similar patches of zenith sky, on the same layer, but in two different directions of the sun's vertical (Karandikar 1955). If the two patches $P_1$ & $P_2$, as shown in fig. 1, are chosen on the sun's meridian and inclined at the same angle to the vertical, then their illumination depend essentially on Sun's depression. At the moment of sunset, i.e. the solar depression D = 0 degree, the two patches are illuminated by the same bile of solar rays. Since $P_1$ & $P_2$ lay on a thin scattering layer at high zenith distance Z from the earth's surface, then their brightness $I_1$ & $I_2$ will be nearly the same. During the course of the twilight, i.e. when the sun's depression increases from D = 0 to $D_1$ the intensity I of the patch $P_1$ is higher than that of the patch $P_2$. So, the difference in the brightness of $P_1$ & $P_2$, i.e., $(I_1 - I_2)$ and the ratio $I_1/I_2$ increases slowly until solar depression $D_1$. For moving the sun from $D_1$ to $D_2$ the brightness $I_1$ decreased much faster than $I_2$; therefore, $(I_1 - I_2)$ and $I_1/I_2$ decreases. The reasons will be given at the results and discussions.

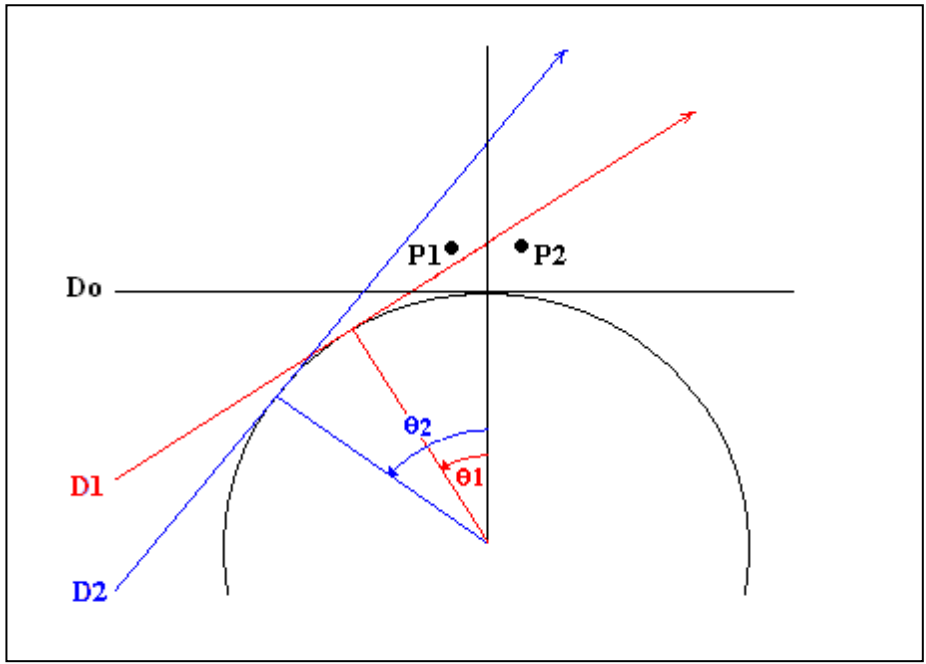

**Fig. 1: Sketch diagram of the earth's shadow at different sun's depressions (D).**

## Results and Discussions

To study the discontinuity in the sky twilight brightness, we have used our data measured at Daraw ($\phi$ = 24° 22 N & $\lambda$ = 32° 58 E) by Asaad et al. (1977), and at Abu-Simbel ($\phi$ = 22° 20 N & $\lambda$ =31° 38′ E) by Nawar (1981, 1983) and given in Tables I and II for Abu-Simbel and Daraw respectively.

The results of the sky twilight brightness measured at the two sites were expressed in number of stars of 10$^{th}$ visual magnitude per square degree for different wavelengths. The effective wavelengths of the filters used are 4400 Å, 5500 Å at Daraw and 4410 Å, 5500 Å & 7900 Å at Abu-Simbel.

It has been found that from the work of Asaad et al. (1977) and Nawar (1981, 1983), there is no evidence for the discontinuity in the slope of the sky twilight brightness curve with sun's depression. The same result has been obtained for all altitudes and all azimuths from direction of sunset as well as for blue, yellow and red colors.

The discontinuity in the sky twilight brightness has been also examined using the two methods mentioned by Karandikar (1955), which discussed before. The discontinuity can be detected by calculating the difference in brightness of two similar patches of the zenith sky situated on solar and ant- solar verticals, at different sun's depression. In the present work, using Tables I and II, the logarithm of difference of twilight brightness [log ($I_1 - I_2$)] at different altitudes in the sky have been calculated for different values of sun's depression and for different colors. The results are drawn in Fig. 2. The figure shows the relation between log ($I_1 - I_2$) and sun's depression D for altitudes 5, 10, 30, 50, 70 degrees at Daraw and for altitudes 10, 30, 50, 70 degrees at Abu- simbel. It can be seen from this figure that there is no evidence of the discontinuity in the sky twilight brightness from the beginning of sunset till 15 degree sun's depression blow the horizon. This result confirms the result obtained by Karandikar (1954). This result is expected, since the difference between $I_1$ & $I_2$ is so big at the same depression angle of the sun. So this method is not applicable to study the discontinuity in the brightness of the sky twilight. Therefore we shall apply the second method. This method depends essentially on the ratio $I_1/I_2$. Using Tables I and II, this ratio has been calculated for different altitudes and sun's depression. The results are drawn in Fig. 3. The figure shows the relation between $I_1/I_2$ and sun's depression for different colors at Daraw and Abu-simbel. It can be seen from this figure that, at altitudes higher than or equal 30 degrees, the ratio $I_1/I_2$ increases with increasing sun's depression and reach it's maximum at sun's depression from 10 to 12 degrees at Daraw, and from 6 to 7 degrees at Abu-simbel. With increasing sun's depression, the ratio $I_1/I_2$ decreases. For altitudes higher than 30 degrees, the ratio $I_1/I_2$ is nearly constant for the sites.

The above mentioned results can be explained using the principle mentioned before at section (2) due to the crossing of the luminescent layers in the upper atmosphere by the earth's shadow. So if we choose two patches $p_1$ and $p_2$ lie on the same altitude, but at large zenith distance, then the two points and all other points in the sky are illuminated by direct sunlight, when the sun is above the horizon. During the course of twilight the earth's shadow rises at the anti solar vertical. With increasing the sun's depression and for altitudes equal or less than 30 degrees, the point $P_2$ lays in earth's shadow, while the other point $P_1$ outside it. So the brightness of $P_2$ suddenly decreases while the brightness of $P_1$ slowly decreases. This gives a sudden increase in the ratio $I_1/I_2$. At higher altitudes the two point $P_1$ and $P_2$ will lie outside the earth's shadow and $I_1/I_2$ still nearly constant. With increasing sun's depression the earth's shadow rises and the point $P_1$ lies in it. A sudden decrease in its brightness and the ratio $I_1/I_2$ gradually decreases with increasing sun's depression.

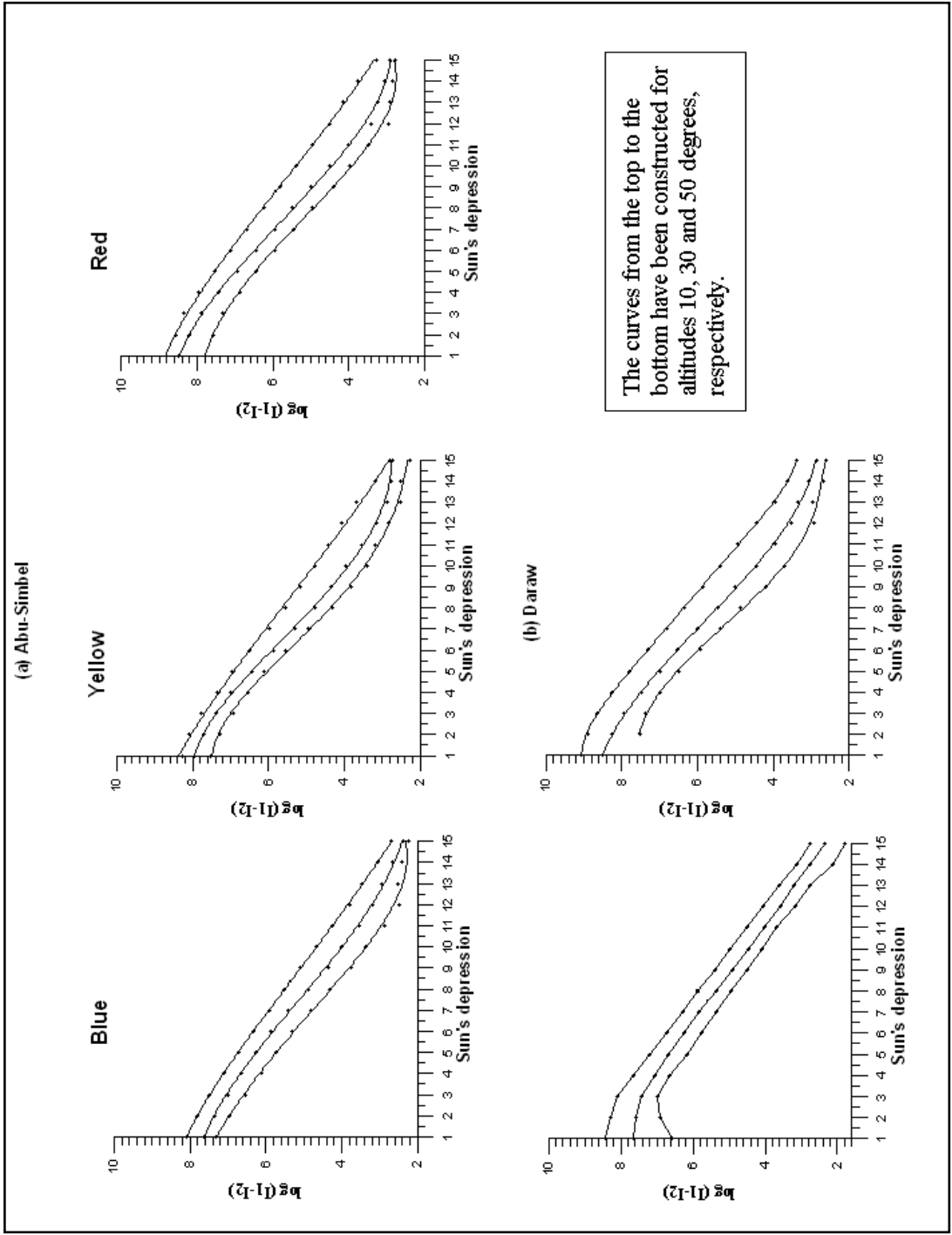

Fig. 2: The relation between log ($I_1-I_2$) and sun's depression at different altitudes and for blue, yellow and red colors.

The curves from the top to the bottom have been constructed for altitudes 10, 30 and 50 degrees, respectively.

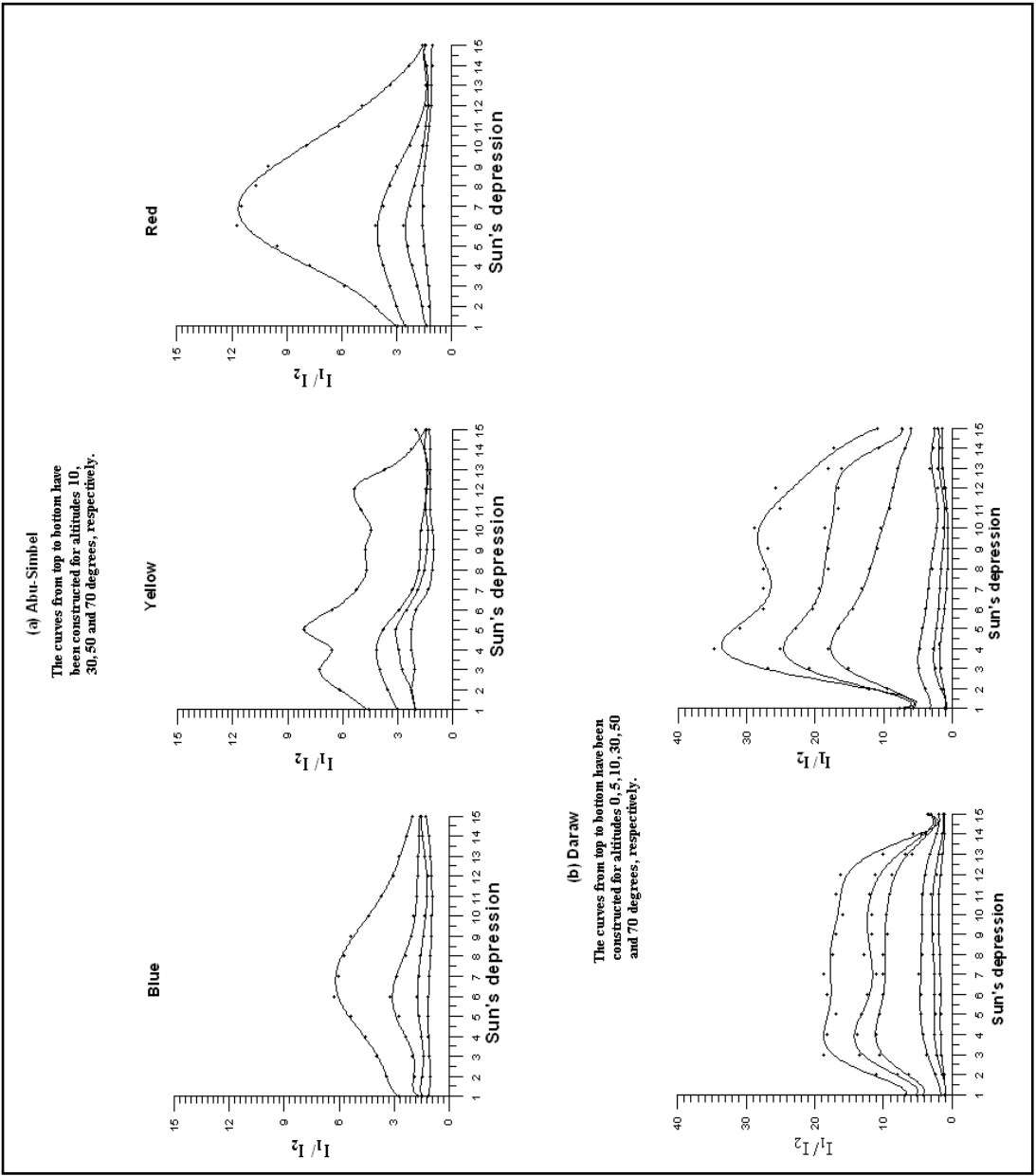

Fig. 3: The relation between log ($I_1/I_2$) and sun's depression at different altitudes and for blue, yellow and red colors.

## Table 1
### Abu Simbel (Blue)

| | Solar vertical | | | | Anti Solar vertical | | | |
|---|---|---|---|---|---|---|---|---|
| D | 10 | 30 | 50 | 70 | 10 | 30 | 50 | 70 |
| 1 | 8.28 | 8.03 | 7.82 | 7.65 | 7.84 | 7.8 | 7.65 | 7.6 |
| 2 | 7.96 | 7.68 | 7.45 | 7.23 | 7.43 | 7.4 | 7.28 | 7.22 |
| 3 | 7.63 | 7.3 | 7.07 | 6.88 | 7.03 | 7 | 6.92 | 6.85 |
| 4 | 7.21 | 6.88 | 6.6 | 6.4 | 6.55 | 6.5 | 6.42 | 6.35 |
| 5 | 6.8 | 6.45 | 6.13 | 5.91 | 6.07 | 6.01 | 5.92 | 5.86 |
| 6 | 6.4 | 6.03 | 5.67 | 5.42 | 5.6 | 5.52 | 5.42 | 5.37 |
| 7 | 6 | 5.59 | 5.22 | 4.97 | 5.22 | 5.13 | 5 | 4.94 |
| 8 | 5.59 | 5.12 | 4.78 | 4.51 | 4.83 | 4.74 | 4.59 | 4.5 |
| 9 | 5.18 | 4.66 | 4.32 | 4.05 | 4.45 | 4.35 | 4.18 | 4.07 |
| 10 | 4.77 | 4.3 | 3.99 | 3.74 | 4.13 | 4.01 | 3.87 | 3.76 |
| 11 | 4.39 | 3.92 | 3.64 | 3.41 | 3.82 | 3.68 | 3.56 | 3.45 |
| 12 | 3.98 | 3.57 | 3.32 | 3.12 | 3.5 | 3.34 | 3.25 | 3.14 |
| 13 | 3.66 | 3.32 | 3.12 | 2.94 | 3.22 | 3.09 | 2.99 | 2.93 |
| 14 | 3.31 | 3.06 | 2.91 | 2.77 | 2.95 | 2.84 | 2.75 | 2.73 |
| 15 | 2.98 | 2.8 | 2.7 | 2.6 | 2.68 | 2.6 | 2.52 | 2.5 |

### Abu Simbel (Red)

| | Solar vertical | | | | Anti Solar vertical | | | |
|---|---|---|---|---|---|---|---|---|
| D | 10 | 30 | 50 | 70 | 10 | 30 | 50 | 70 |
| 1 | 8.98 | 8.7 | 8.34 | 8.05 | 8.51 | 8.3 | 8.2 | 7.98 |
| 2 | 8.69 | 8.38 | 7.99 | 7.72 | 8.07 | 7.9 | 7.78 | 7.63 |
| 3 | 8.42 | 8.05 | 7.65 | 7.38 | 7.65 | 7.52 | 7.37 | 7.3 |
| 4 | 8.01 | 7.56 | 7.16 | 6.9 | 7.12 | 6.99 | 6.83 | 6.76 |
| 5 | 7.58 | 7.07 | 6.67 | 6.42 | 6.6 | 6.47 | 6.29 | 6.24 |
| 6 | 7.15 | 6.57 | 6.18 | 5.93 | 6.08 | 5.95 | 5.76 | 5.72 |
| 7 | 6.73 | 6.1 | 5.71 | 5.45 | 5.67 | 5.53 | 5.35 | 5.26 |
| 8 | 6.29 | 5.65 | 5.25 | 4.99 | 5.26 | 5.12 | 4.94 | 4.79 |
| 9 | 5.85 | 5.18 | 4.77 | 4.5 | 4.85 | 4.7 | 4.52 | 4.33 |
| 10 | 5.44 | 4.75 | 4.4 | 4.17 | 4.54 | 4.39 | 4.2 | 4.04 |
| 11 | 5.03 | 4.35 | 4.03 | 3.84 | 4.24 | 4.08 | 3.88 | 3.75 |
| 12 | 4.62 | 3.92 | 3.66 | 3.5 | 3.93 | 3.76 | 3.56 | 3.46 |
| 13 | 4.32 | 3.76 | 3.53 | 3.37 | 3.79 | 3.6 | 3.41 | 3.32 |
| 14 | 4.01 | 3.59 | 3.4 | 3.21 | 3.64 | 3.43 | 3.26 | 3.18 |
| 15 | 3.7 | 3.43 | 3.27 | 3.08 | 3.5 | 3.27 | 3.1 | 3.05 |

## Abu Simbel (Yellow)

| | Solar vertical | | | | Anti Solar vertical | | | |
|---|---|---|---|---|---|---|---|---|
| D | 10 | 30 | 50 | 70 | 10 | 30 | 50 | 70 |
| 1 | 8.36 | 8.11 | 7.82 | 7.62 | 7.7 | 7.63 | 7.5 | 7.32 |
| 2 | 8.16 | 7.85 | 7.52 | 7.27 | 7.37 | 7.3 | 7.17 | 6.92 |
| 3 | 7.84 | 7.52 | 7.12 | 6.87 | 6.98 | 6.92 | 6.68 | 6.55 |
| 4 | 7.42 | 7.12 | 6.72 | 6.47 | 6.6 | 6.5 | 6.25 | 6.12 |
| 5 | 7.03 | 6.58 | 6.27 | 6 | 6.12 | 6 | 5.78 | 5.65 |
| 6 | 6.57 | 6.04 | 5.77 | 5.52 | 5.75 | 5.57 | 5.37 | 5.22 |
| 7 | 6.07 | 5.57 | 5.26 | 4.97 | 5.35 | 5.23 | 4.98 | 4.85 |
| 8 | 5.65 | 5.12 | 4.76 | 4.52 | 4.98 | 4.85 | 4.57 | 4.47 |
| 9 | 5.25 | 4.72 | 4.35 | 4.11 | 4.57 | 4.47 | 4.2 | 4.1 |
| 10 | 4.87 | 4.35 | 3.98 | 3.76 | 4.22 | 4.11 | 3.85 | 3.72 |
| 11 | 4.5 | 3.98 | 3.67 | 3.47 | 3.8 | 3.79 | 3.5 | 3.38 |
| 12 | 4.15 | 3.68 | 3.37 | 3.22 | 3.42 | 3.52 | 3.22 | 3.14 |
| 13 | 3.8 | 3.42 | 3.15 | 2.98 | 3.23 | 3.27 | 3.03 | 2.9 |
| 14 | 3.45 | 3.22 | 3 | 2.8 | 3.1 | 3.03 | 2.82 | 2.72 |
| 15 | 3.22 | 3.1 | 2.82 | 2.73 | 3.05 | 2.8 | 2.67 | 2.62 |

## Table 2
## Daraw (Blue)

| | Solar vertical | | | | | | Anti Solar vertical | | | | | |
|---|---|---|---|---|---|---|---|---|---|---|---|---|
| D | 0 | 5 | 10 | 30 | 50 | 70 | 0 | 5 | 10 | 30 | 50 | 70 |
| 1 | 8.83 | 8.68 | 8.56 | 8.04 | 7.76 | 7.59 | 8 | 7.96 | 7.94 | 7.82 | 7.73 | 7.64 |
| 2 | 8.67 | 8.49 | 8.36 | 7.8 | 7.48 | 7.33 | 7.63 | 7.59 | 7.56 | 7.42 | 7.34 | 7.26 |
| 3 | 8.45 | 8.27 | 8.12 | 7.58 | 7.23 | 7.02 | 7.18 | 7.14 | 7.1 | 7.01 | 6.88 | 6.82 |
| 4 | 8 | 7.84 | 7.7 | 7.18 | 6.85 | 6.58 | 6.74 | 6.7 | 6.66 | 6.56 | 6.44 | 6.37 |
| 5 | 7.53 | 7.38 | 7.24 | 6.8 | 6.41 | 6.16 | 6.3 | 6.26 | 6.22 | 6.14 | 6.02 | 5.94 |
| 6 | 7.13 | 6.92 | 6.78 | 6.36 | 5.98 | 5.73 | 5.87 | 5.83 | 5.78 | 5.7 | 5.58 | 5.5 |
| 7 | 6.69 | 6.42 | 6.34 | 5.92 | 5.56 | 5.3 | 5.42 | 5.38 | 5.34 | 5.24 | 5.14 | 5.06 |
| 8 | 6.22 | 6.06 | 5.9 | 5.46 | 5.14 | 4.86 | 4.98 | 4.95 | 4.9 | 4.82 | 4.7 | 4.62 |
| 9 | 5.8 | 5.6 | 5.45 | 5.02 | 4.7 | 4.46 | 4.57 | 4.53 | 4.48 | 4.38 | 4.26 | 4.18 |
| 10 | 5.36 | 5.18 | 5.02 | 4.58 | 4.28 | 4.02 | 4.16 | 4.11 | 4.04 | 3.94 | 3.82 | 3.73 |
| 11 | 4.9 | 4.72 | 4.56 | 4.14 | 3.86 | 3.58 | 3.67 | 3.64 | 3.6 | 3.5 | 3.38 | 3.28 |
| 12 | 4.46 | 4.27 | 4.12 | 3.72 | 3.42 | 3.16 | 3.25 | 3.22 | 3.18 | 3.13 | 3.06 | 2.96 |
| 13 | 4 | 3.8 | 3.68 | 3.36 | 3.08 | 2.82 | 3 | 2.97 | 2.92 | 2.86 | 2.78 | 2.68 |
| 14 | 3.5 | 3.38 | 3.26 | 3 | 2.68 | 2.52 | 2.75 | 2.72 | 2.68 | 2.64 | 2.54 | 2.46 |
| 15 | 3.06 | 2.99 | 2.92 | 2.66 | 2.42 | 2.26 | 2.5 | 2.47 | 2.44 | 2.38 | 2.3 | 2.22 |

## Daraw (Yellow)

| D | Solar vertical | | | | | | Anti Solar vertical | | | | | |
|---|---|---|---|---|---|---|---|---|---|---|---|---|
|   | 0 | 5 | 10 | 30 | 50 | 70 | 0 | 5 | 10 | 30 | 50 | 70 |
| 1 | 9.49 | 9.33 | 9.16 | 8.66 | 8.18 | 7.94 | 8.6 | 8.49 | 8.4 | 8.14 | 8.22 | 7.88 |
| 2 | 9.29 | 9.1 | 8.94 | 8.38 | 7.9 | 7.68 | 8.2 | 8.02 | 7.96 | 7.78 | 7.66 | 7.56 |
| 3 | 9.05 | 8.88 | 8.68 | 8.04 | 7.6 | 7.32 | 7.62 | 7.56 | 7.5 | 7.34 | 7.2 | 7.08 |
| 4 | 8.64 | 8.46 | 8.28 | 7.58 | 7.2 | 6.9 | 7.1 | 7.06 | 7.02 | 6.9 | 6.76 | 6.62 |
| 5 | 8.17 | 8 | 7.82 | 7.1 | 6.72 | 6.38 | 6.68 | 6.64 | 6.6 | 6.46 | 6.34 | 6.2 |
| 6 | 7.72 | 7.54 | 7.34 | 6.64 | 6.23 | 5.84 | 6.28 | 6.23 | 6.18 | 6.04 | 5.92 | 5.78 |
| 7 | 7.28 | 7.08 | 6.86 | 6.16 | 5.74 | 5.3 | 5.84 | 5.79 | 5.74 | 5.61 | 5.48 | 5.34 |
| 8 | 6.83 | 6.64 | 6.39 | 5.66 | 5.28 | 4.8 | 5.39 | 5.38 | 5.31 | 5.18 | 5.06 | 4.92 |
| 9 | 6.4 | 6.18 | 5.92 | 5.2 | 4.76 | 4.3 | 4.97 | 4.92 | 4.88 | 4.74 | 4.62 | 4.48 |
| 10 | 5.96 | 5.74 | 5.46 | 4.7 | 4.32 | 3.88 | 4.5 | 4.47 | 4.44 | 4.32 | 4.2 | 4.02 |
| 11 | 5.5 | 5.28 | 4.98 | 4.23 | 3.8 | 3.52 | 4.1 | 4.06 | 4.02 | 3.88 | 3.76 | 3.62 |
| 12 | 5.05 | 4.83 | 4.52 | 3.8 | 3.48 | 3.2 | 3.64 | 3.61 | 3.58 | 3.46 | 3.34 | 3.18 |
| 13 | 4.46 | 4.38 | 4.02 | 3.52 | 3.24 | 2.98 | 3.2 | 3.17 | 3.12 | 3 | 2.9 | 2.8 |
| 14 | 4.16 | 3.92 | 3.7 | 3.26 | 3.02 | 2.8 | 2.92 | 2.89 | 2.86 | 2.8 | 2.74 | 2.64 |
| 15 | 3.8 | 3.6 | 3.48 | 3.08 | 2.9 | 2.7 | 2.76 | 2.73 | 2.7 | 2.66 | 2.58 | 2.5 |